\documentclass{TTP_DSL2006}

\usepackage[dvipsone]{graphicx}
\usepackage[intlimits]{amsmath}
\usepackage{amssymb}
\usepackage{exscale}
\begin{document}

\title{ Canted spiral magnetic order in layered systems}

%\author{ FULL First Author \inst{1}$^{,\rm{a}}$ \textbf{,} FULL Second Author \inst{2}$^{,\rm{b}}$ and Last Author \inst{3}$^{,\rm{c}}$}
\author{ TIMIRGAZIN Marat $^{\rm{a}}$\text{,} GILMUTDINOV Vitaly
$^{\rm{b}}$\text{,} and ARZHNIKOV Anatoly $^{\rm{c}}$}

\institute{ Physical-Technical Institute, Ural Branch of Russian Academy of Sciences, Kirova str. 132, Izhevsk
426000, Russia}

\maketitle

\vspace{-3mm}
\sffamily

\begin{center}
$^{\rm a}$timirgazin@gmail.com, $^{\rm b}$vitaliodestroyer@gmail.com, $^{\rm c}$arzhnikof@bk.ru
\end{center}

\vspace{2mm} \hspace{-7.7mm} \normalsize \textbf{Keywords:} Hubbard model, spiral magnetic structure, incommensurate
magnetism, layered systems. \\

\vspace{-2mm} \hspace{-7.7mm}
\rmfamily

\noindent \textbf{Abstract.} Formation of a canted spiral magnetic order is studied in the framework of a mean-field
approximation of the
Hubbard model. It is revealed that this magnetic state can be stabilized under certain conditions in layered systems
with a
relatively small interplane electron hopping. Example of an experimentally observed magnetic structure of
La$_{2-x}$Sr$_x$CuO$_4$ is considered. It is shown that the canting magnetic order can be described in terms of a simple
non-relativistic band magnetism.

\section{Introduction}
\vspace{-6pt}

It is generally accepted that interplay between charge carriers and magnetic correlations is responsible for
high-temperature superconductivity in cuprates. Magnetic structure forms an environment for hole motion, and
determination of its characteristics and of its evolution with doping is therefore very important problem in clarifying
the mechanism of superconductivity.

Parent CuO$_2$-based compounds, e.g. La$_2$CuO$_4$, are quasi-two-dimensional Heisenberg antiferromagnets. With doping
of Sr atoms, the magnetic structure of La$_2$CuO$_4$ significantly changes. Presence of additional holes
favors incommensurate spin density waves formation. Neutron scattering in
La$_{2-x}$Sr$_x$CuO$_4$ reveals coexistence of both commensurate and incommensurate magnetic structures in the vicinity
of half-filling (hole doping $x<0.02$) \cite{La1}.
At $x\sim 0.02$ the system goes to the incommensurate (spin glass) state with the magnetic  structure wave vector
$\mathbf{Q}=(Q,Q)$. For $x>0.06$  a magnetic structure with the wave vector $\mathbf{Q}=(Q,\pi)$ is stabilized
\cite{La2}.

Theoretical calculations based on Hartree-Fock and slave-boson approaches in the Hubbard model confirm the formation
of incommensurate (spiral) magnetic structures with the hole doping \cite{Sarker91,Fresard92,Igoshev10,Igoshev13}. Phase
diagrams of the 2D Hubbard model for the relation of next-nearest neighbor hopping to nearest neighbor hopping
$t'/t=0.2$, which approximately corresponds to La$_{2-x}$Sr$_x$CuO$_4$ \cite{Hybertsen92}, qualitatively fully
reproduce the magnetic phase transitions observed in experiment with doping \cite{Igoshev10,Igoshev13}. 

\begin{figure}
 \includegraphics[width=1\textwidth]{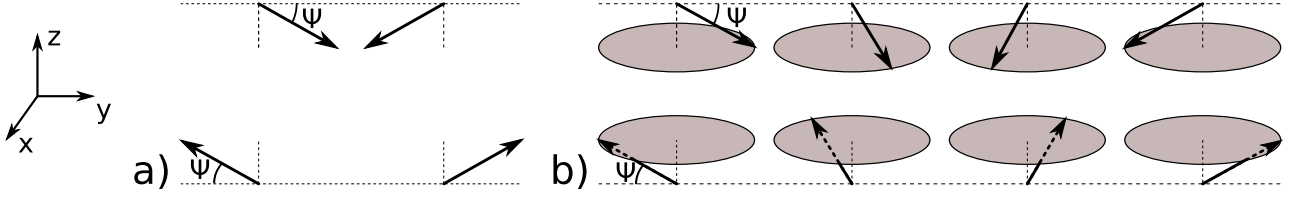}
 \caption{Sketch of a 3D canted spin order: a) antiferromagnetic in La$_2$CuO$_4$ b) spiral in La$_{2-x}$Sr$_x$CuO$_4$.
The canting angle $\psi$ from the $xy$ CuO planes is exaggerated.}
\end{figure}

The in-plane
antiferromagnetic exchange interaction between Cu atoms is typically $J_{||}\sim0.1eV$, as was determined by inelastic
neutron scattering \cite{Birgeneau89} and by Raman scattering. Weak interaction between the CuO$_2$ layers
(interplane exchange interaction is estimated to be $J_\bot\sim10^{-5}J_{||}$ \cite{Cheong88}) gives
rise to the three-dimensional long-range N\'{e}el order ($T_N\approx325$K) \cite{Keimer92,Kastner98}. Additionally, each
CuO plane has a
weak perpendicular magnetic moment, and the interplane exchange orients these moments
antiferromagnetically as shown in Fig. 1a. The spins, hence, are canted 0.17$^{\circ}$ away from
$xy$-plane \cite{Thio88}. The canting is shown to persist with a light doping of Sr ($x=0-0.03$) \cite{Lavrov01}.
The authors of \cite{Lavrov01} note an unusual susceptibility anisotropy observed in doped region, which can be
explained
by a spin-density wave magnetic structure. The resulting supposed magnetic structure of La$_{2-x}$Sr$_x$CuO$_4$ can
then be presented as a combination of the spiral type of spin-density wave with a canting (Fig. 1b.).

The canting is traditionally associated with a Dzyaloshinskii-Moriya interaction which is appeared due to orthorhombic
structure in La$_2$CuO$_4$. In the current work we study another scenario of such magnetic structure formation, which
is based only on simple Hubbard model assumptions and does not require spin-orbit coupling. For this we use the 3D
Hubbard model taking into account spiral magnetic states, with the interplane electron hopping much smaller than the
in-plane hopping.

\section{Formalism}

The Hubbard Hamiltonian of the considered layered system is divided into the kinetic energy of electrons hopping
between sites $K$, and the repulsive Coulomb on-site interaction $V$:
\begin{eqnarray} \label{Hamiltonian}
 \mathcal{H}=K+V=\sum_{\langle\alpha,\beta\rangle ,\langle i,j\rangle ,\sigma}t^{}_{\alpha,\beta, i,j}c^\dag_{\alpha,
i,\sigma}c^{}_{\beta,j,\sigma}+U\sum_{\alpha, j}n_{\alpha, j,\uparrow}n_{\alpha, j,\downarrow},
\end{eqnarray}
where $\langle\alpha,\beta\rangle=1,2$ are numbers of planes, $\langle i,j \rangle$ are in-plane site numbers, $\sigma$
is spin number.

We suppose the interplane transfer integral $t_z$ to be much smaller than the in-plane transfer integral $t_{xy}$. After
Fourier transformations $K$ takes the form:
\begin{eqnarray}
 K=\sum_{\alpha, \mathbf{k}, \sigma} \varepsilon_{\mathbf{k}}^{xy}c^\dag_{\alpha, \mathbf{k},\sigma}c^{}_{\alpha,
\mathbf{k},\sigma} + \sum_{\mathbf{k},\sigma}
\varepsilon_\mathbf{k}^z(c^\dag_{1,\mathbf{k},\sigma}c^{}_{2,\mathbf{k},\sigma}+c^\dag_{2,\mathbf{k},\sigma}c^{}_{1,
\mathbf { k }, \sigma }),
\end{eqnarray}
where $\varepsilon_\mathbf{k}^{xy}=-2t_{xy}(\cos{k_x}+\cos{k_y})+4t'_{xy}\cos{k_x}\cos{k_y}$ and
$\varepsilon_\mathbf{k}^{z}=-2t_z\cos{k_z}$ 
are the in-plane and interplane dispersion laws. Nearest and next-nearest neighbors electron hopping is taken
into account in plane.

Interaction term $V$ can be written in terms of electron and spin densities using
$n_{\uparrow}n_{\downarrow}=n^2/4-\mathbf{S}^2$. Then in the mean-field (Hartree-Fock) approximation:
\begin{eqnarray}
 V=UN\left(M^2-\frac{n^2}{4}\right)+2U\sum_{\alpha, j} \left(\frac{n}{4} n_{\alpha, j}-\mathbf{M}_{\alpha,
j}\mathbf{S}_{\alpha, j}\right),
\end{eqnarray}
where $n=\langle n_{\alpha,i}\rangle$ is a uniform average electron concentration, $\mathbf{M}_{\alpha,
j}=\langle\mathbf{S}_{\alpha,j}\rangle$ is average magnetization, $M$ is its amplitude, $N$ is full number of sites in
the
system.

The magnetic structure depicted on Fig.1b is described by following vector:
\begin{eqnarray}
 \mathbf{M_{\alpha,j}}=(M\cos{\mathbf{QR}_j}\cos\psi,M\sin{\mathbf{QR}_j}\cos\psi,M(-1)^{\alpha+1}\sin\psi),
\end{eqnarray}
with $\mathbf{Q}$ being the in-plane wave vector. Taking this into account, we obtain:
\begin{eqnarray}
 V=UN\left(M^2-\frac{n^2}{4}\right)+\frac{Un}{2}\sum_{\alpha, j} n_{\alpha, j} - \nonumber \\
 -UM \sum_{\alpha, j} \left(
S^+_{\alpha j}e^{-i\mathbf{QR}_j}\cos\psi+S^-_{\alpha j}e^{i\mathbf{QR}_j}\cos\psi+2S^z_{\alpha
j}(-1)^{\alpha+1}\sin\psi \right)
\end{eqnarray}

In $k$-space, we have:
\begin{eqnarray}
 V=UN\left(M^2-\frac{n^2}{4}\right)+\frac{Un}{2}\sum_{\alpha, \mathbf{k}, \sigma} c^\dag_{\alpha, \mathbf{k},
\sigma}c^{}_{\alpha, \mathbf{k}, \sigma} - UM \sum_{\alpha,
\mathbf{k}} \big( c^\dag_{\alpha, \mathbf{k}, \uparrow}c^{}_{\alpha, \mathbf{k+Q}, \downarrow}\cos\psi + \nonumber \\
+ c^\dag_{\alpha, \mathbf{k+Q}, \downarrow}c^{}_{\alpha, \mathbf{k}, \uparrow}\cos\psi +  
(c^\dag_{\alpha, \mathbf{k}, \uparrow}c^{}_{\alpha, \mathbf{k}, \uparrow}-c^\dag_{\alpha, \mathbf{k+Q},
\downarrow}c^{}_{\alpha, \mathbf{k+Q}, \downarrow})(-1)^{\alpha+1}\sin\psi \big).
\end{eqnarray}

Full Hamiltonian (\ref{Hamiltonian}) represents a quadratic form of creation and annihilation operators
$c_{1,\mathbf{k},\uparrow}$,
$c_{1,\mathbf{k+Q},\downarrow}$, $c_{2,\mathbf{k},\uparrow}$, $c_{2,\mathbf{k+Q},\downarrow}$. To diagonalize the
Hamiltonian, one
must solve a polynomial equation of 4-th degree. We have used the Ferrari's method for this and have found eigenvalues
$\varepsilon_{\mathbf{k},l}$ and eigenvectors $T_{\mathbf{k},lm}$ in each $k$-point. The electron concentration $n$,
magnetic moment $M$ and total energy $E$ can be expressed as:
\begin{eqnarray}
 n=\frac{1}{N} \sum_{\mathbf{k}, l}f(\varepsilon_{\mathbf{k}, l}), \\
 M=\frac{1}{N\cos{\psi}}\sum_{\mathbf{k},
l}(T_{\mathbf{k},1l}^*T_{\mathbf{k},2l}+T^*_{\mathbf{k},3l}T_{\mathbf{k},4l})f(\varepsilon_{\mathbf{k},l}), \\
 \label{energy}
 E=UN\left(M^2-\frac{n^2}{4}\right)+ \sum_{\mathbf{k}, l}\varepsilon_{\mathbf{k}, l}f(\varepsilon_{\mathbf{k}, l}),
\end{eqnarray}
where $f(\varepsilon_{\mathbf{k} l})$ is the Fermi function.

In order to obtain the ground state of the system for given parameters $n,U,t_{xy},t'_{xy},t_z$ the total energy should
be minimized with respect to all $\mathbf{Q}$ and $\psi$.

\section{Results}

The magnetic phase diagram of the ground state of the 2D Hubbard model in Hartree-Fock approximation for $t'=0.2$, which
approximately corresponds to real La$_{2-x}$Sr$_x$CuO$_4$ value, was calculated in \cite{Igoshev10} (Fig. 3a in
reference). Our investigation is focused on hole-doped side of this diagram which contains $(Q,Q)$ and $(Q,\pi)$ spiral
magnetic, ferromagnetic, and antiferromagnetic phases. The values of $U/t$ studied, approximately correspond
to realistic values of $U/t$ for cuprates which lie in the range $8-20$ according to the ab initio calculations
\cite{Hybertsen90}. The value of $t_z$ is taken to be $0.1t_{xy}$.

\begin{figure}[!h]
  \hskip 0.2\textwidth
 \includegraphics[width=0.5\textwidth]{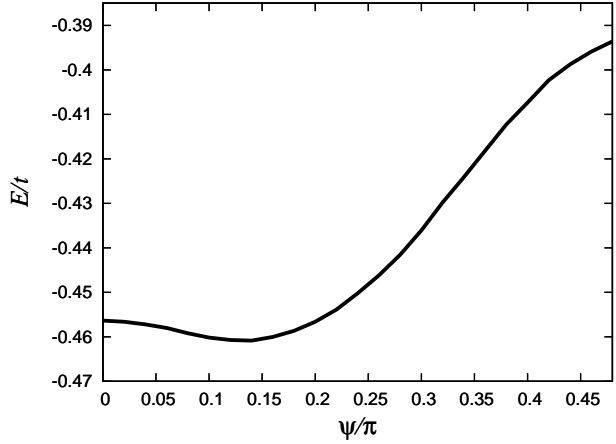}
 \caption{Dependence of the total energy $E$ upon the canting angle $\psi$ for $U/t_{xy}=10, n=0.9, t'_{xy}=0.2,
t_z=0.1, \mathbf{Q}=(0.65\pi,\pi)$.}
\end{figure}

We considered both $(Q,Q)$ and $(Q,\pi)$ spiral states to determine
if the addition of the weak interplane electron hopping can stabilize the canted spiral magnetic order. Technically,
this is reduced to finding (for some parameters $U/t,n$) a $Q$ value that minimizes the energy (\ref{energy}) with
$\psi=0$ for each direction of $\mathbf{Q}$ vector, and next to finding a $\psi$ angle that minimizes the energy
(\ref{energy}) for this $Q$.

Our investigation shows that the canting is not energetically favorable for the $(Q,Q)$ state for all the parameters
studied. But for the $(Q,\pi)$ phase this is not the case. In Fig. 2 the dependence of the total energy $E$ upon the
canting angle $\psi$ for $U/t_{xy}=10, n=0.9, \mathbf{Q}=(0.65\pi,\pi)$ is presented. The minimum of energy
corresponds to $\psi=0.14\pi$, which means that the canted spiral state is the most energetically favorable among the
$(Q,\pi)$ states at these parameters. It should be noted that this is not the ground state because the phase separation
between $AF$ and $FM$ phases is still more favorable, as in the 2D system \cite{Igoshev10}.

Our result can be qualitatively associated with the results of paper \cite{Chubukov95}, where stability of $(Q,Q)$
and $(Q,\pi)$ phases in respect to transverse spin fluctuations was studied for small $t'/t$ and hole doping. It was
found that from these two states only $(Q,\pi)$ is unstable and tends to form a non-coplanar spin configuration.

The results obtained show a possibility of stabilization of the canted spiral magnetic state in the systems with
weakly coupled layers in the simple Hubbard model without account of spin-orbit interaction. So, we have
demonstrated an alternative non-relativistic scenario of formation of the canted magnetic structure which is not based
on the Dzyaloshinskii-Moriya interaction.

\section{Acknowledgements}

This work was supported by the Russian Foundation for Basic Research (projects Nos. 14-02-31603-mol\_a, 12-02-00632-a),
by the Ural Branch of Russian Academy of Sciences (No. 14-2-NP-273), by the Presidium of the Russian Academy of Sciences
(No. 12-U-2-1021).

\end{document}